\documentclass{article}
\usepackage{moreverb}
\usepackage{tikz}
\usepackage[numbers]{natbib}
\usepackage{booktabs}
\usepackage{amsmath}
\usepackage{amsfonts}
\usepackage{authblk}
\usepackage{algorithm}
\usetikzlibrary{arrows.meta}
\usepackage{algorithm}
\usepackage{algpseudocode}
\usepackage{graphicx} %% for Jpegs\usetikzlibrary{arrows.meta}s
\usetikzlibrary{shapes,positioning}
 \usepackage{url}
\usepackage{moreverb}

\newcommand\BibTeX{{\rmfamily B\kern-.05em \textsc{i\kern-.025em b}\kern-.08em
T\kern-.1667em\lower.7ex\hbox{E}\kern-.125emX}}

\usepackage{subcaption}
\usepackage{enumitem}
\usepackage{comment}

\setlist[description]{font=\normalfont\itshape\textbullet\space}

\begin{document}

\title{Assessment of evidence against homogeneity in exhaustive subgroup treatment effect plots}

\author[1]{Björn Bornkamp}
\author[2]{Jiarui Lu}
\author[1,3]{Frank Bretz}
\affil[1]{Advanced Methodology and Data Science, Novartis Pharma AG, Basel, Switzerland}
\affil[2]{Department of Biometrics, Vertex Pharmaceuticals Inc., Boston, MA, USA}
\affil[3]{Institute of Medical Statistics, Center for Medical Data Science, Medical University of Vienna, Vienna, Austria}

\maketitle

\abstract{Exhaustive subgroup treatment effect plots display treatment effect estimates for all subgroups of interest against subgroup sample size, providing a useful overview of observed treatment effect heterogeneity in a clinical trial. As in any exploratory subgroup analysis, however, the estimates may be affected by small sample sizes and multiplicity, making it difficult to judge whether the observed heterogeneity is noteworthy or simply due to chance. To facilitate more interpretable exploratory assessments, we introduce a computationally efficient method for assessing evidence against homogeneity in exhaustive subgroup treatment effect plots. Using the Doubly Robust (DR) learner, we construct pseudo-outcomes to estimate subgroup effects and derive reference distributions that quantify how surprising the observed heterogeneity is under an assumption of homogeneous treatment effects. We introduce and compare three methods for computing probabilities from the reference distribution: numerical integration, a Bonferroni approximation, and a wild bootstrap approach. The resulting evidence is summarized using surprise values and used to categorize subgroups in the plots according to the strength of evidence against homogeneity. The methods are illustrated using data mimicking a cardiovascular trial and evaluated in a simulation study, showing well-calibrated inference and improved performance over standard approaches based on simple differences of observed group means.
}


\renewcommand\thefootnote{}

\renewcommand\thefootnote{\fnsymbol{footnote}}
\setcounter{footnote}{1}

\section{Introduction}
\label{sec:introduction}

Subgroup analyses present significant challenges in biostatistics, particularly in clinical trials. Estimating treatment effects within subgroups is often unreliable due to limited sample sizes and multiplicity issues. This frequently leads to findings that fail to generalize beyond the observed data, as is often seen when initial subgroup findings are not replicated in subsequent trials \citep{Yusuf1991,bretz2014multiplicity,wallach2017evaluation,ruberg2021assessing}. These challenges primarily affect exploratory subgroup analyses, whereas pre-specified analyses with adequate sample sizes and multiple testing strategies are generally more reliable \citep{alosh2017tutorial}.

In any clinical trial, some treatment effect heterogeneity will be observed. The key question is whether the observed heterogeneity is greater than would be expected under a scenario of homogeneous treatment effects. One way to assess this is to investigate how likely the observed subgroup differences are under a model assuming homogeneous treatment effects. This can be achieved with a global interaction test, and several authors have emphasized its importance for evaluating subgroup treatment effect heterogeneity in both univariate and multivariate settings \citep{chernozhukov2017generic,Schandelmaier:2020,kent2020a,watson2020machine,harrell:2025, ding2016randomization, dukes2024nonparametric}. Although interaction tests are widely used, clinical trials are not typically powered for them, as they are not of primary importance. Therefore, the resulting p-values should be interpreted with caution and not as strict binary decision rules.

Recently, Muysers and others\citep{muysers2020systematic, kirsch2022subscreen} advocated an exhaustive display of subgroup treatment effects. Rather than focusing on a limited set of subgroups, as in a forest plot, they proposed plotting treatment effect estimates for all single- and multi-level overlapping subgroups of interest against their respective sample sizes. Their aim was to efficiently calculate point estimates for multiple subgroups to explore those that may differ from the overall treatment effect. We refer to this display as exhaustive subgroup treatment effect plots, with subgroup size on the x-axis and subgroup treatment effect on the y-axis (see Figure~\ref{fig:funnel_wo_lines} for an example). This graphical approach provides context by comparing subgroups of similar size, offering a comprehensive assessment rather than selective reporting. Importantly, the authors deliberately avoided inferential statistics such as p-values or confidence intervals, to prevent distracting attention from clinically meaningful subgroups. Instead, they recommend using these plots as a basis for a research team discussion to assess treatment effect homogeneity or heterogeneity.

To further aid interpretation, we develop a computationally efficient method to assess the observed differences between subgroup treatment effects and the overall treatment effect under a model of homogeneous treatment effects. This allows us to highlight specific subgroups whose observed treatment effects are surprising under the  model of homogeneous treatment effects. More specifically, we apply the Doubly Robust (DR) learner\citep{kennedy2023towards} to construct pseudo-outcomes for the individual treatment effect of each patient, estimate subgroup treatment effects by averaging these pseudo-outcomes within subgroups, and compute probabilities and thresholds from the underlying reference distribution. We interpret the resulting p-values not as binary decision rules, but as divergence p-values\citep{Greenland2023}. These quantify the degree of incompatibility of the observed data with a model of homogeneous treatment effects \citep{cole2020}. Small p-values indicate greater incompatibility with the homogeneity model, whereas large p-values indicate less incompatibility.

The remainder of this paper is structured as follows. Section~\ref{sec:example} introduces a synthetic simulated data set mimicking data from a cardiovascular outcome trial to motivate our approach. Section~\ref{sec:meth} presents the proposed methodology. Section~\ref{sec:simulation} evaluates its operating characteristics through simulation. Section~\ref{sec:example_rev} revisits the motivating example, and Section~\ref{sec:discussion} concludes.

\section{Case study}
\label{sec:example}

To illustrate the methods in this article, we use synthetic simulated data that mimic data from a cardiovascular outcome
trial comparing three investigational doses against placebo. For the purpose of this paper, we focus on the comparison of the highest dose group against placebo for an exploratory outcome of interest, high-sensitivity C-reactive protein (hs-CRP), an inflammation marker that is measured three months post baseline. The dataset includes 3297 patients in the placebo group and 2194 in the highest dose group. Note that the sample size of this trial was determined for the primary time-to-event endpoint. The power to detect differences in hs-CRP is considerably higher than would typically be the case in analyses of treatment effect heterogeneity targeting the primary endpoint. This setting therefore provides greater opportunity to detect differences in treatment effects across subgroups compared to standard situations where treatment effect heterogeneity is explored for a primary endpoint.

For reasons of confidentiality, we use simulated data. To estimate the overall treatment effect we analyze the change from baseline in log-transformed hs-CRP values using a linear model adjusting for baseline log10-transformed hs-CRP, focusing on the highest dose versus placebo. This analysis yields a treatment effect estimate of $-0.38$ with a 95\% confidence interval of $(-0.40, -0.36)$, corresponding to a reduction of approximately 59\% in hs-CRP relative to baseline compared with placebo. Thus, a clear treatment effect is present in the data for this outcome.

In addition, we employ the exhaustive subgroup treatment effect plot\citep{muysers2020systematic}. As potential effect modifiers, we consider 19 baseline variables described in Table \ref{tbl:variables} in Appendix \ref{sec:app_bl}, most of which were pre-specified for subgroup analyses of the primary endpoint in the original study. Numerical baseline variables were categorized into tertiles. We consider all single-variable subgroups as well as two-variable cross-classifications, restricting to subgroups with at least 50 patients in both treatment and control arms. This results in a total of 1408 subgroups.

\begin{figure}
    \centering
	\includegraphics[width=6in]{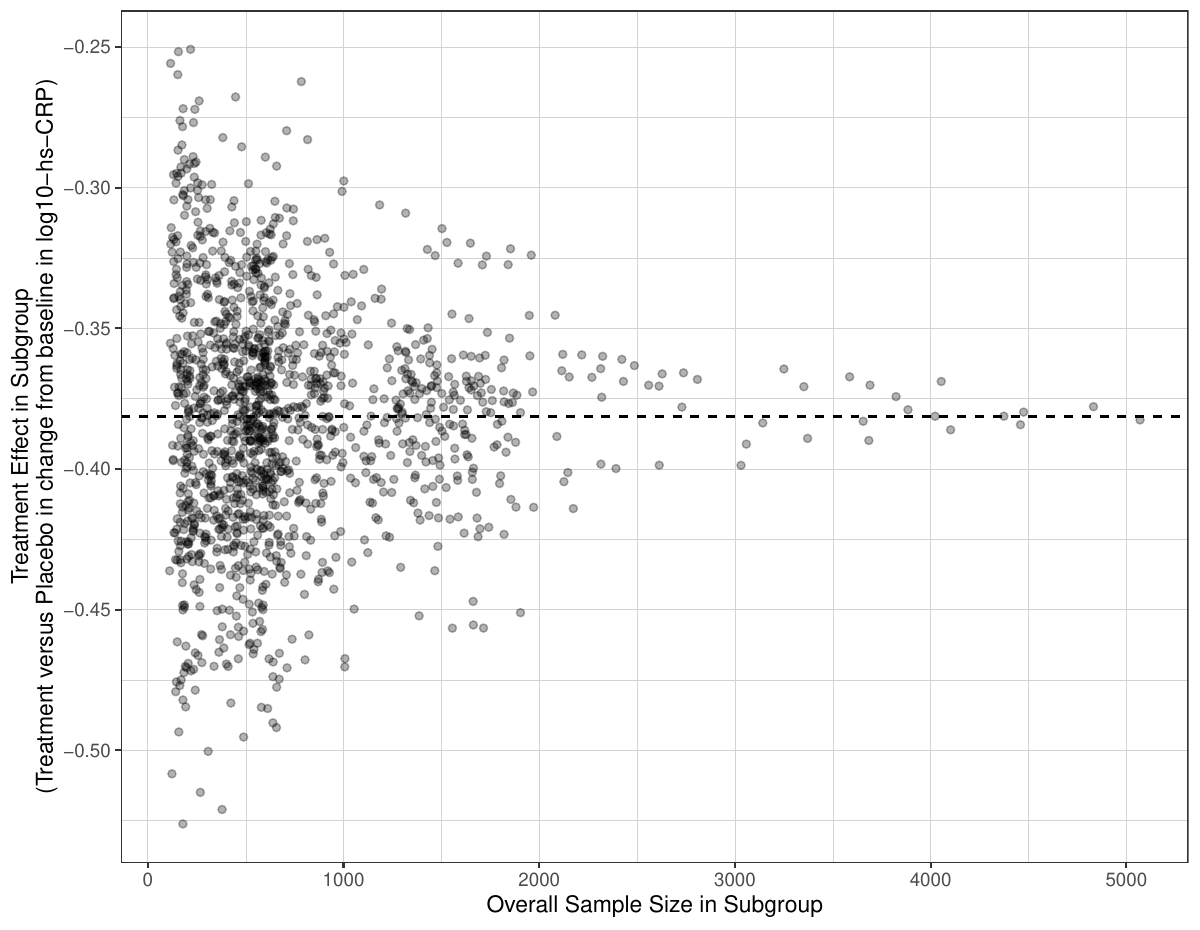} 
    \caption{Exhaustive subgroup treatment effect plot, showing 1408 one-level or two-level subgroup treatment effects. The dashed line corresponds to the overall treatment effect estimate.}
    \label{fig:funnel_wo_lines}
\end{figure}

Figure~\ref{fig:funnel_wo_lines} displays the treatment effect estimates of all these subgroups on the y-axis against subgroup size on the x-axis. As expected, the plot exhibits a funnel shape: smaller subgroups show greater variability in treatment effect estimates, whereas larger subgroups display reduced variability and converge toward the overall treatment effect estimate, reflecting greater overlap with the overall population. The R package \emph{subscreen}\citep{muysers2020systematic, kirsch2022subscreen} provides an interactive version of such a plot, allowing users either to identify points corresponding to specific subgroups of interest or to click on points to reveal the associated subgroup. Such a display provides a useful overview of how treatment effects vary with baseline covariates.

As discussed in Section~\ref{sec:introduction}, the original approach does not provide a formal method to assess evidence against homogeneity. Nevertheless, we believe that such exploratory investigations can be further enhanced by considering how surprising the observed heterogeneity of certain subgroups is under a model of homogeneous treatment effects. Along the lines of Questions 3 and 4 posed by Stephen Senn\citep{senn2004controversies}, research teams may naturally ask the following questions based on an exhaustive subgroup treatment effect plot: 
\begin{itemize}
    \item[(A)] How heterogeneous are the treatment effects across all subgroups? To what extent is the observed heterogeneity consistent with the hypothesis that all subgroups share the same treatment effect?
    \item[(B)]  Which subgroups exhibit larger or smaller treatment effects compared with the overall trial population?
\end{itemize} 
We return to these questions in Section~\ref{sec:example_rev}.

\section{Methodology}
\label{sec:meth}

Creating an exhaustive subgroup treatment effect plot requires calculating subgroup treatment effects for a potentially large number of subgroups, which can be computationally intensive if statistical models are fit separately for each subgroup. To address this challenge, we propose using pseudo-outcomes $\varphi_i$ for the individual treatment effect of each patient $i$, based on a two-stage doubly robust estimator, known as the DR learner \citep{kennedy2023towards}. Subgroup treatment effects can then be estimated by averaging the pseudo-outcomes $\varphi_i$ within each subgroup, reducing computational complexity from (potentially iterative) model fitting to simple averaging.

Assume observed data $Z_i=(Y_i, A_i, X_i), i\in \{1,\ldots,N\}$, where $Y \in \mathbb{R}$ denotes the outcome of interest, $A \in \{0,1\}$ a binary treatment indicator, and $X \in \mathbb{R}^d$ the covariates. We condition on the realized baseline covariates $X_1,\ldots,X_N$. Conditional on $X_1,\ldots,X_N$, we assume that the pairs $Y_i,A_i$, $i=1,\ldots,N$, are independent across patients, with $Y_i$ allowed to depend on both $A_i$ and $X_i$. The collection of subgroups may be defined as functions of the baseline covariates. Consequently, after conditioning on $X_1,\ldots,X_N$, the subgroup indicators, subgroup memberships, and subgroup sizes are treated as fixed. This includes subgroup definitions based on baseline covariate cutpoints determined from the realized covariate distribution (cutpoints informed by outcome data are not considered in this paper).  
%For the rest of the manuscript, the randomness is from $Y_i$ and $A_i$, unless otherwise specified. For simplicity, we use $\mathbb{E}(Y | X_i=x_i)$ and $\mathbb{E}(Y | X_i)$ interchangeably since $X_i$ is considered as fixed and known value.

The intuition behind pseudo-outcomes in the DR learner is to construct a new outcome that combines information from both a propensity score model $\pi(x) = \mathbb{P}(A=1 | X=x)$ (the probability of receiving treatment given the covariates) and an outcome model $\mu_a(x) = E(Y | X=x, A=a)$ (the expected outcome for a patient conditional on covariates and treatment). Assuming for now that $\pi(x), \mu_0(x)$, and $\mu_1(x)$ are known (we consider estimation of these quantities later in Section \ref{ssec:nuisance}), the pseudo-outcomes are constructed as \citep{kennedy2023towards}
\begin{equation}
    \label{eq:pseudo_outcome}
    \varphi(Z_i) = \mu_1(X_i) - \mu_0(X_i) + \frac{A_i - \pi(X_i)}{\pi(X_i)(1 - \pi(X_i))}  \left(Y_i - \mu_{A_i}(X_i) \right).
\end{equation}
By construction, $E(\varphi(Z_i) | X_i) = \mu_1(X_i)-\mu_0(X_i)$, so that the conditional expectation of the pseudo-outcome equals the conditional treatment effect. Under the conditions stated in Hahn\cite{hahn1998role} and Kennedy\cite{kennedy2024semiparametric}, averaging the pseudo-outcomes $\varphi(Z_i)$ yields a semiparametrically efficient estimator of the average treatment effect for the overall population. Similarly, the most efficient effect estimate for a given subgroup is the average of the $\varphi(Z_i)$ within that subgroup. In what follows we omit the argument $Z_i$ and simply write $\varphi_i$ for $\varphi(Z_i)$.

Next, consider index sets $\mathcal{S}_j$ containing the patients in subgroup $j \in \{1, \ldots, k\}$, where $k$ denotes the total number of subgroups. Because we condition on the realized baseline covariates, the sets $\mathcal{S}_j$ and their sizes $N_j = |\mathcal{S}_j|$ are treated as fixed. Note that the subgroups will generally overlap. Let $\widehat{\delta}_{j}=\frac{1}{N_j}\sum_{i\in \mathcal{S}_j}\varphi_i$ be an estimator of the treatment effect in subgroup $j$ and $\widehat{\delta}=\frac{1}{N}\sum_i \varphi_i$ be an estimator of the overall treatment effect, where $N=|\bigcup_j \mathcal{S}_j|$.
Motivated by Questions~(A) and~(B) in Section~\ref{sec:example}, we assess the $k$ estimated treatment effect differences
$\widehat{\Delta}_{j} = \widehat{\delta}_{j} - \widehat{\delta}$ and determine whether the magnitude of $\widehat{\Delta}_{j}$ is large relative to its variability under the assumption of no subgroup treatment effect heterogeneity.

To facilitate derivations, we re-write $\widehat{\Delta}_{j}=\sum_{i=1}^Nw_{i,j}\varphi_i$ with $w_{i,j}=\frac{\mathbf{1}(i\in \mathcal{S}_j)}{N_j}-\frac{1}{N}$, where $\mathbf{1}(i\in \mathcal{S}_j)=1$ if patient $i$ is in $\mathcal{S}_j$ and 0 otherwise. By conditional independence across patients of the $\varphi_i$ we obtain the variance 
\begin{equation}
\label{eqn:variance}
    V_j=Var(\widehat{\Delta}_j|X_1,...,X_N)=\sum_{i=1}^Nw_{i,j}^2Var(\varphi_i|X_i)
\end{equation} and the covariance 
\begin{equation}
    \label{eqn:covariance}
Cov(\widehat{\Delta}_j,\widehat{\Delta}_l|X_1,...,X_N)=\sum_{i=1}^Nw_{i,j}w_{i,l}Var(\varphi_i|X_i).
\end{equation}

We aim to assess the consistency of the data with the homogeneity assumption that $E(\widehat{\Delta}_j|X_1,\ldots,X_N)=0$ for all $j$, that is, that the average treatment effect in each investigated subgroup equals the overall treatment effect (Question A). Under this assumption, the standardized test statistics $T_j={\widehat{\Delta}_{j}}\big/{\sqrt{V_{j}}}$ are centered at zero, conditional on the realized covariates, for $j=1,\ldots, k$. To assess the overall evidence against homogeneity, we consider the maximum statistic $T_{\max}=\max_{j}  \left|T_j\right|$ and compute the divergence p-value $p=1-F\big(T_{\max}\big)$, where $F$ denotes the reference distribution function of $T_{\max}$ under the homogeneity assumption above. We interpret this p-value as a measure of the degree of incompatibility between the observed data and what would be expected under a model of no subgroup treatment effect heterogeneity \citep{Greenland2023}: a small $p$ indicates greater incompatibility with the homogeneity assumption used to derive the reference distribution. We emphasize that this assessment is conditional on the realized covariates and applies to the specific collection of subgroups under study. Accordingly, $F$ is the conditional reference distribution of the maximum standardized deviation over this finite family of overlapping subgroups.

Additionally, we can assess which subgroups show evidence of a differential treatment effect compared to the overall population (Question B) by computing the individual divergence p-values $p_j=1 - F(|T_j|)$, with $F$ as defined above. These p-values are therefore calibrated against the maximum statistic rather than the marginal reference distribution of each $T_j$. For the subgroup with the maximum absolute test statistic, the corresponding p-value coincides with the global p-value for     assessing treatment effect homogeneity. For the remaining subgroups, the $p_j$ can be interpreted as multiplicity adjusted measures of incompatibility with the homogeneity model over the finite family of subgroups under study. The direction of the observed departure, that is, whether the subgroup treatment effect is larger or smaller than the overall treatment effect, is determined by the sign of $\widehat{\Delta}_j$. %Under the model of no treatment effect heterogeneity, the differences between subgroup treatment effects and the overall treatment effect will all fall within the $\gamma$-homogeneity region with probability $\gamma$. In other words, whenever $T_{\max} > q_{\gamma}$, at least one subgroup will fall outside the $\gamma$-homogeneity region.

We now discuss three approaches for computing probabilities from the reference distribution $F$. For a fixed finite collection of subgroups, under a multivariate Lindeberg-Feller condition\cite{billingsley1995probability}, each $T_j$ is asymptotically standard normal and the joint distribution of $\{ T_1, \ldots, T_k\}$ is asymptotically multivariate normal, where the correlation matrix is obtained from the variances and covariances in \eqref{eqn:variance} and \eqref{eqn:covariance}. Accordingly, the corresponding probabilities can be calculated using numerical integration as implemented in, for example, the \emph{mvtnorm} package in R \citep{genz:bret:2009}.  If the  total number of subgroups $k$ is large (say in the thousands), numerical integration may be infeasible, as the required matrix computations for the growing correlation matrix become very time-consuming. In such cases, one may rely on approximations to the multivariate normal probability integral that consist of combinations of lower dimensional problems based on approximations to Boole’s formula. The most widely known of these approximations is the inequality by Bonferroni\citep{bonferroni1936teoria}. 

Alternatively, we can approximate the reference distribution $F$ by a wild bootstrap \citep{mammen1993bootstrap} under the homogeneous effect assumption. Let $\hat{e}_i=\varphi_i-\widehat{\delta}$ denote residuals for the pseudo outcomes under a homogeneity assumption. Following Algorithm~\ref{alg:wild_bootstrap}, we draw for $b=1,\ldots,N_{boot}$ independent Rademacher variables $g_{i,b}$, for all observations $i$, which take the values $-1$ and $1$ with probability $0.5$ each, and compute
\begin{equation*}
T_{j,b}=\sum_{i=1}^N w_{i,j}\,(\widehat{\delta}+\hat{e}_i g_{i,b})\Big/\sqrt{V_j}, \qquad T_{\max,b}=\max_{j}\left|T_{j,b}\right|.
\end{equation*}
Each bootstrap sample keeps the residual of every patient and only flips it randomly around the overall treatment effect. So the resampled test statistics all have mean 0, consistent with treatment effect homogeneity. More specifically, the wild bootstrap operates under the stronger reference model that the conditional treatment effect is assumed to be constant and equal to the overall treatment effect for every patient, this implies $E(\widehat{\Delta}_j\mid X_1,\ldots,X_N)=0$ for every investigated subgroup $j$. Note that the $V_j$ are calculated once and are fixed in the bootstrap resamples. The covariance between two subgroup treatment differences is $\sum_{i} w_{i,j}w_{i,l}\hat{e}_i^2$ in the bootstrap resamples, the empirical counterpart of \eqref{eqn:covariance}. So the bootstrap distribution reproduces the heteroskedastic dependence between the subgroup test statistics (even when the variance $V_j$ used to standardize the $j$th bootstrap test statistic is not correct). The p-values can be estimated via $p=\frac{1+|\{b: T_{\max,b}\geq T_{\max}\}|}{N_{boot}+1}$ and $p_j=\frac{1+|\{b: T_{\max,b}\geq |T_j|\}|}{N_{boot}+1}$. The justification of the wild bootstrap is asymptotic. Theoretical properties have been investigated also for the case of maximum type statistics\citep{chernozhukov2013gaussian}.

%Alternatively, we can approximate the reference distribution $F$ using a wild bootstrap approach. 
%Specifically, we permute the $\varphi_i$ values while keeping subgroup indicators fixed. For each of $N_{perm}$ permutations, we compute $T_{j,\ell}$ for subgroup $j$, and then calculate the maximum statistic $T_{\max,\ell} = \max_{j}  \left|T_{j,\ell}\right|$, $\ell = 1, \ldots, N_{perm}$. The resulting permutation distribution $\{T_{\max,\ell}\}$ asymptotically approximates the sampling distribution of $T_{\max}$. 

%The quantity $q_{\gamma}=F^{-1}(\gamma)$ used in defining the $\gamma$-homogeneity region can then be estimated as the $\gamma$-quantile of $\{T_{\max,\ell}\}$, $\ell = 1, \ldots, N_{perm}$. 

\begin{algorithm}
\caption{Wild bootstrap approximation of the reference distribution}
\textbf{Input}: Number of bootstrap samples $N_{boot}$; $k$ investigated subgroups represented by index sets $\mathcal{S}_j$, $j \in \{1,\ldots,k\}$; pseudo-outcomes $\varphi_i$, $i\in\{1,\ldots,N\}$; variances $V_j$
\begin{algorithmic}[1]
\State Calculate $\widehat{\delta}=\frac{1}{N}\sum_{i=1}^N \varphi_i$ and residuals $\widehat{e}_i=\varphi_i-\widehat{\delta}$.
\For{$b \in \{1,\ldots,N_{boot}\}$}
    \State Generate i.i.d. $g_{i,b}\in\{-1,1\}$ with equal probability, $i\in\{1,\ldots,N\}$.
    \For{$j \in \{1,\ldots,k\}$}
        \State Calculate 
        $T_{j,b}=
        \frac{\sum_{i=1}^N w_{i,j}\widehat{e}_i g_{i,b}}
        {\sqrt{V_j}}$,
        where $w_{i,j}=\frac{\mathbf{1}(i\in \mathcal{S}_j)}{N_j}-\frac{1}{N}$.
    \EndFor
    \State Calculate $T_{\max,b}=\max_{j\in\{1,\ldots,k\}}|T_{j,b}|$.
\EndFor
\State Approximate $F$ by the empirical distribution $\widehat{F}$ of $\{T_{\max,b}\}_{b=1}^{N_{boot}}$.
\State Estimate the $\gamma$-quantile $q_\gamma$ of $F$ by $\widehat{F}^{-1}(\gamma)$.
\end{algorithmic}
\label{alg:wild_bootstrap}
\end{algorithm}

One might question the advantage of using individual pseudo-outcomes, $\varphi_i$, for estimating subgroup treatment effects rather than relying on simple differences in group means across subgroups. A key limitation of using raw group means, particularly for small subgroups, is that chance imbalances in prognostic baseline covariates between treatment groups can lead to less precise subgroup-specific treatment effect estimates. In contrast, the pseudo-outcomes $\varphi_i$ incorporate information from baseline covariates through the outcome regression and can therefore be viewed as a form of covariate adjustment that potentially improves the precision and efficiency of subgroup treatment effect estimation. We evaluate this potential efficiency gain in the simulation study in Section~\ref{sec:simulation}.

\iffalse % commented out
\begin{algorithm}
\caption{Permutation approach for computing probabilities for the reference distribution $F$}\label{alg: perm}
\begin{algorithmic}
\State \textbf{Input} Number of permutations $N_{perm}$; $k$ pre-specified subgroups represented by index sets $\mathcal{S}_j$, $j \in \{1,\ldots,k\}$; pseudo-outcomes $\varphi_i$; variance $\sigma^2$.
\State \textbf{Output} Empirical distribution of the maximum statistic $T_{\max}$. 
\State \textbf{for} $\ell = 1, \ldots, N_{perm}$ \textbf{do}
\State \indent  Step 1: Permute the $\varphi_i$.
\State \indent \textbf{for} $j = 1, \ldots, k$ \textbf{do}
\State \indent \hspace{\parindent} Step 2: Compute $T_{j,\ell}$ based on subgroup $j$ and the $\ell$-th permuted data.
\State \indent \textbf{end for} 
\State \indent  Step 3: Compute $T_{\max,\ell} = \max_{j}  \left|T_{j,\ell}\right|$ for the $\ell$-th permuted data.
\State \textbf{end for} 
\State \textbf{Step 4}: Estimate the $\gamma$-quantile of the reference distribution $F$ of the maximum statistic $T_{\max}$, $q_{\gamma}$, by taking the $\gamma$-quantile of 
$\{T_{\max,\ell}\}$, $\ell = 1, \ldots, N_{perm}$. 
\end{algorithmic}
\end{algorithm}
\fi % end commented out

\subsection{Special case: homogeneous variance of the pseudo-outcomes}
\label{ssec:homoskedastic}

It is instructive to consider the case when the variance $Var(\varphi_i|X_i)=\sigma^2$ is constant. Then \eqref{eqn:variance} reduces to $\sigma^2({N_j}^{-1} - {N}^{-1})$, and it becomes transparent that when the subgroup size $N_j$ approaches $N$, the variance of $\widehat{\Delta}_j$ approaches 0 as the subgroup increasingly overlaps with the overall population. Conversely, smaller subgroups have greater variability, giving rise to the characteristic funnel shape described below when subgroup treatment effect estimates are plotted against subgroup size.

Under the same homoskedasticity assumption, the covariance between two subgroup contrasts has a simple closed form expression. Let
$N_{j\ell}=|\mathcal{S}_j \cap \mathcal{S}_\ell|$ denote the number of subjects belonging to both subgroups $j$ and $\ell$. Then
\eqref{eqn:covariance} reduces to
$$
Cov(\widehat{\Delta}_j,\widehat{\Delta}_\ell | X_1,...,X_N)=\sigma^2\left(\dfrac{N_{j\ell}}{N_jN_\ell} - \dfrac{1}{N}\right).
$$
Consequently, the corresponding correlation is
\begin{align}
  \rho_{j\ell} = \dfrac{\dfrac{N_{j\ell}}{N_jN_\ell}  - \dfrac{1}{N}}{\sqrt{\left(\dfrac{1}{N_j} - \dfrac{1}{N}\right)\left(\dfrac{1}{N_\ell} - \dfrac{1}{N}\right)}}.   \nonumber
\end{align}

Thus, under homoskedasticity, the correlation between the subgroup contrasts is determined entirely by the subgroup sizes and their degree of overlap. Strongly overlapping subgroups will generally be positively correlated. Even non-overlapping subgroups are generally not independent: if $N_{j\ell}=0$, their covariance is $-\sigma^2/N$, reflecting the fact that both subgroup effects are compared with the same estimated overall treatment effect. From the formula one can also see that the correlation of a subgroup to its complement is $-1$.

The simple variance expression also allows the reference distribution to be represented directly in the exhaustive subgroup treatment effect plot. Let $T_j={\widehat{\Delta}_{j}}\big/{\sigma\sqrt{{N_j}^{-1} - {N}^{-1}}}$, $T_{\max}=\max_{j=1, \ldots,k}  \left|T_j\right|$ and let $q_{\gamma}$ denote the $\gamma$-quantile of the reference distribution of $T_{\max}$ under the assumption of treatment effect homogeneity. The condition $\left|T_j\right| < q_{\gamma}$ is equivalent to $\widehat{\delta} - q_{\gamma}\sigma\sqrt{{N_j}^{-1} - {N}^{-1}} \leq \widehat{\delta}_j \leq \widehat{\delta} + q_{\gamma}\sigma\sqrt{{N_j}^{-1} - {N}^{-1}}$. Hence, we can construct $\gamma$-homogeneity regions for the subgroup treatment effects by plotting 
\begin{equation}
    \label{eq:homo_region}
\widehat{\delta} \pm q_{\gamma}\sigma\sqrt{{N_j}^{-1} - {N}^{-1}}
\end{equation} 
as a function of subgroup size $N_j$. Equivalently, the corresponding homogeneity region for the treatment effect difference $\widehat{\Delta}_j$ is centered at zero with limits $\pm q_\gamma\sigma\sqrt{N_j^{-1}-N^{-1}}$. In practice, $\sigma$ is replaced by an estimate, for example the empirical standard deviation of the pseudo-outcomes under the homoskedastic working model.

%The $\gamma$-homogeneity regions provide a direct graphical representation of the reference distribution. Their width decreases monotonically with subgroup size and approaches zero as $N_j$ approaches $N$. Importantly, under homoskedasticity the width depends on the subgroup only through $N_j$. Consequently, all subgroups of the same size share the same homogeneity limits, producing smooth funnel-shaped bands around the estimated overall treatment effect. 
%Under the model of no treatment effect heterogeneity, the differences between subgroup treatment effects and the overall treatment effect will all fall within the $\gamma$-homogeneity region with probability $\gamma$. In other words, whenever $T_{\max} > q_{\gamma}$, at least one subgroup will fall outside the $\gamma$-homogeneity region.

Note that homoskedasticity of the pseudo-outcomes is an additional assumption and does not follow from treatment effect homogeneity. The general variance and covariance expressions in \eqref{eqn:variance} and \eqref{eqn:covariance} therefore remain the primary focus. In that setting, when the conditional variances differ across patients, the width of the corresponding reference region depends not only on subgroup size but also on the composition of the subgroup, and two subgroups with the same $N_j$ need not have the same variance.

\subsection{Estimation of nuisance parameters}
\label{ssec:nuisance}

\begin{algorithm}
\caption{Calculation of pseudo-outcomes}
\textbf{Input}: $\boldsymbol{Z} = \{\{Y_i, A_i, \mathbf{X}_i\}|\,i\in \{1,...,N\}\}$
\begin{algorithmic}[1]
\State Split the indices $i=\{1,...,N\}$ into $\kappa$ random subsets (folds), $\Theta_1,...,\Theta_\kappa$ stratified by treatment group.
\For{$k \in \{1, \ldots, \kappa\}$}
    \State Let $\Theta_{train,k} = \{1,...,N\} \setminus \Theta_k$
    \State Estimate nuisance functions: $\widehat{\pi}, \widehat{\mu}_0$ and $\widehat{\mu}_1$ using data in $\Theta_{train,k}$
    \State For all $i \in \Theta_{k}$ calculate pseudo-outcomes: $\widehat{\varphi}_i = \widehat{\mu}_1 (\mathbf{X}_i) - \widehat{\mu}_0 (\mathbf{X}_i)+ \frac{A_i - \widehat{\pi}(\mathbf{X}_i)}{\widehat{\pi}(\mathbf{X}_i)\left(1-\widehat{\pi}(\mathbf{X}_i)\right)} \left(Y_i - \widehat{\mu}_{A_i} (\mathbf{X}_i) \right) $ 
\EndFor 
\end{algorithmic}
\label{alg:pseudo_outcomes}
\end{algorithm}

So far we have assumed the pseudo-outcomes $\varphi_i$ and their conditional variances $Var(\varphi_i|X_i)$ to be known. In practice, both must be estimated. We construct the estimated pseudo-outcomes $\widehat{\varphi}_i$ using the cross-fitting procedure in Algorithm~\ref{alg:pseudo_outcomes}, slightly modified from Kennedy\citep{kennedy2023towards}, following Sechidis and others\citep{sechidis2025using}: the propensity and outcome models are fitted on part of the data and then applied to the remaining, held-out patients, so that each patient's pseudo-outcome is computed from models that were not fitted on that patient. 

In a randomized clinical trial, our main setting, the randomization probability is known. Since the outcome models used for a patient were fitted on other patients, and treatment was assigned at random, each $\widehat{\varphi}_i$ is unbiased for the treatment effect of a patient with covariates $X_i$, even if the outcome model is misspecified (see Appendix \ref{sec:app_dr}). Hence, under homogeneity, the estimated subgroup differences remain centered at zero. In addition, if we assume that the fitted outcome models stabilize as the sample size grows, that is, they converge to some fixed (not necessarily the true) functions, the estimated pseudo-outcomes then asymptotically behave like independent pseudo-outcomes, so that the results of Section~\ref{sec:meth} carry over. A misspecified outcome model hence does not invalidate the procedure, but only increases the variance of the pseudo-outcomes and thus reduces the ability to detect heterogeneity. If the propensity score has to be estimated, as in observational studies, the double robustness of the pseudo-outcomes ensures consistent subgroup effect estimates if either the propensity score or the outcome models are estimated consistently. For valid inference, however, both need to be estimated consistently, and the product of their estimation errors must vanish sufficiently fast\citep{kennedy2023towards}.

%A concern is whether replacing these unknown quantities by estimates distorts the reference distribution used for inference. Here two situations can be distinguished. In a randomized clinical trial, which is our main setting, the treatment-assignment probabilities are known by design and don't need to be estimated. In this case each estimated pseudo-outcome is unbiased, equal to the treatment effect for patients with the same baseline characteristics, regardless of whether the outcome models are misspecified. This follows from the so-called "double robustness property" of the pseudo-outcome $\varphi_i$. 
%The outcome model therefore does not determine whether the procedure is valid; it affects only how precisely the subgroup effects are estimated, and hence how much evidence against homogeneity the method can detect. 
%Consequently, under the assumption of homogeneous treatment effects the subgroup differences remain centered at zero, and the reference distribution is unaffected by the choice of outcome model. When the propensity score is not known and must itself be estimated, as in observational studies, both models must be estimated consistently and the product of their estimation errors must vanish faster than the sampling variability of the subgroup effects. Because it is the product of the two errors that matters, this is not an overly strong condition.

The estimation of the variance $V_j$ additionally requires replacing $Var(\varphi_i|X_i)$ with an estimate. Loosely motivated by heteroskedastic variance estimation techniques in linear regression, we propose to use $(\widehat{\varphi}_i-\widehat{\delta})^2$, that is, the squared residual under assumed treatment effect homogeneity. 
%This is a plug-in estimator and its asymptotic property will depend on the asymptotic property of $\widehat{\varphi}_i$ and $\widehat{\delta}$. 
In a randomized clinical trial setting, this quantity
%$(\widehat{\varphi}_i -\widehat{\delta})^2$ will be unbiased under the null model, but 
may overestimate the variance under treatment effect heterogeneity, although in realistic heterogeneous scenarios the systematic component in this residual may be small relative to the error term in the pseudo-outcomes. We evaluate the performance of this estimate in the simulation study in Section~\ref{sec:simulation}.

\section{Simulation study}
\label{sec:simulation}

This section evaluates the proposed approaches through a simulation study. We describe the aims of the study, the data-generating mechanisms with their assumptions and scenarios, the methods being compared, the performance metrics used to evaluate statistical operating characteristics, and a graphical summary of the statistical performance of the methods, based on the simulation results.

\subsection{Aims}
\label{sec:aims}

The aim of the simulation study is to compare the repeated sampling properties of the approaches proposed in Section \ref{sec:meth} for calculating p-values under homogeneous and heterogeneous effect scenarios, and to contrast it with a simpler approach that does not use pseudo-outcomes.

\subsection{Data-generating mechanisms}
\label{sec:datagen}

We adopt the simulation setup implemented in the R package \emph{benchtm} \citep{benchtm, sun2024comparing, chen2025comparing}. We assume a two-arm randomized trial with $N=500$ patients. The treatment indicator $A_i$ for patient $i$ is generated by permuting a vector with 250 zeros and 250 ones. The mean response is
\begin{equation}
\label{form:data-generation}
f(A_i, X_i) = f_{prog}(X_i) + A_i(\beta_0 + \beta_1 f_{pred}(X_i)),
\end{equation}
where $f_{prog}(X_i)$ represents the prognostic component and $f_{pred}(X_i)$ the predictive component, based on baseline covariates $X_i$. For the purpose of this simulation study, we restrict attention to continuous response variables, with $Y_i = f(X_i, A_i) + \epsilon_i$, where $\epsilon_i \sim N(0, 1)$. The covariates $X_i$ are 30-dimensional and generated to mimic a realistic scenario \cite{sun2024comparing}. Table \ref{tab:sim_models} summarizes the four data-generating models considered. These models include step-like (Scenario 1) and linear (Scenario 2) functions, as well as predictive effects defined by multiple covariates (Scenarios 3 and 4).

\begin{table}
\centering
\caption{Data-generating models. Here, $a\lor b$ denotes `a or b', $a\land b$ denotes `a and b', $I(.)$ is the indicator function, $\Phi(.)$ is the cumulative distribution function of the standard normal distribution, and $s$ is a scaling factor depending on the data generating model; see the \emph{benchtm} package for details.}
\label{tab:sim_models}
 \begin{tabular}{l l} 
 \hline
Scenario & Model equation \\ \hline
1 & $f(X_i, A_i) =  s\times (0.5I(X_{1,i}=\text{Y})+X_{11,i}) + A(\beta_0 + \beta_1\Phi(20(X_{11,i}-0.5)))$ \\
2 & $f(X_i, A_i) = s\times (X_{14,i}-I(X_{8,i}=\text{N})) + A(\beta_0 + \beta_1X_{14,i})$ \\ 
3 & $f(X_i, A_i) = s\times (I(X_{1,i}=\text{N})-0.5X_{17,i}) + A(\beta_0 + \beta_1I((X_{14,i}>0.25)\land (X_{1,i}=\text{N})))$ \\ 
4 & $f(X_i, A_i) = s\times (X_{11,i}-X_{14,i}) + A(\beta_0 + \beta_1I((X_{14,i}>0.3)\lor(X_{4,i}=\text{Y})))$ \\ \hline
\end{tabular}
\end{table}

The parameters $\beta_0$ and $\beta_1$ are jointly chosen to achieve a power of 0.5 for an unadjusted Z-test of the overall treatment effect. This reflects a realistic clinical trial setting, where the observed effect is smaller than anticipated at the design stage. In such situations, there may be interest in exploring covariates that modify the treatment effect and identifying subgroups with potentially larger effects.

We determined the value $\beta^*_1$ to ensure 80\% power for the interaction test of $H_0: \beta_1=0$ versus $H_1: \beta_1 \neq 0$ under the true model \eqref{form:data-generation}, at a significance level of 0.1. In our simulation study, we consider $\beta_1=0$ and $\beta_1=2\beta^*_1$ to represent scenarios of treatment effect homogeneity and heterogeneity, respectively. For $\beta_1=2\beta^*_1$, there is thus a high probability of detecting effect modification if $f_{prog}(X)$ and $f_{pred}(X)$ were known.

To generate candidate subgroups based on the 30 baseline covariates $X_i$, numerical variables are categorized into three groups using tertiles of their distributions, while categorical variables define subgroups by each category. In addition, these univariate subgroups are combined with an `and' operator. Subgroups with fewer than 10 patients in either treatment arm are excluded. This leads to approximately 3200–3300 subgroups, depending on the simulated covariates. The numerical integration approach based on the multivariate normal distribution described in Section~\ref{sec:meth} is computationally slow in this dimensionality. We therefore also consider a stricter requirement of at least 60 patients per treatment arm in each subgroup, reducing the number of investigated subgroups to 190–230. Note that the subgroup definitions are not informed by the true predictive models. For instance, the ideal subgroup in Scenario 1 would involve splitting $X_{11}$ at 0.5, whereas in Scenario 4 it would combine $X_4$ and $X_{14}$ with an `or' operator. Since all methods are equally affected, this does not limit the conclusions of this simulation study; rather, it reflects how such analyses would be conducted in practice.

The data generating mechanisms above assume a homoskedastic error variance, which will also imply pseudo-outcomes that are approximately homoskedastic based on Equation \eqref{eq:pseudo_outcome}. We hence also consider a heteroskedastic scenario: For that we use the mean from Scenario 1 in Table \ref{tab:sim_models} and consider $\epsilon_i\sim N(0,\sigma_i^2)$ with $\sigma_i= 0.7\exp(2\log(20)(X_{11} - 0.5))$. As $X_{11}\in[0,1]$ this implies $\sigma_i\in [0.035,14]$. For the given distribution of $X_{11}$ in \textit{benchtm} the implied mean of $\sigma_i$ is given by $1.02$ and 10\%,50\% and 90\% quantiles are given by $0.11$, $0.56$, $2.21$, corresponding to relatively strong heteroskedasticity.

\subsection{Methods}
\label{sec:methods}

We compare the three methods described in Section~\ref{sec:meth} to compute p-values of the maximum statistic $T_{\max}$: numerical integration, Bonferroni approximation, and the wild bootstrap approach. All three methods rely on the pseudo-outcomes $\varphi_i$, estimated using a five-fold cross-fitting procedure as outlined in Algorithm \ref{alg:pseudo_outcomes}. For estimation of the mean functions $\mu_0$ and $\mu_1$ in each fold, we use an ensemble method implemented in the R package \emph{SuperLearner} \citep{polley:2024}, with LASSO and random forest as base learners. In our simulation setting, the LASSO model is correctly specified for $\mu_0$, but for $\mu_1$ only the random forest can reasonably approximate the true functions. For the propensity score model, we use the known randomization probabilities.

The wild bootstrap approach is conducted with $N_{boot}=500$ bootstrap samples. For the multivariate normal integration approach, we use the R package \emph{mvtnorm} with default control parameters. The correlation matrix of the test statistics (Section~\ref{sec:meth}) is positive semidefinite but singular as the test statistics of a subgroup and its complement have a correlation of $-1$. Because rounding errors can then make the matrix numerically indefinite, we use the \emph{Matrix::nearPD} function as a safeguard to compute the nearest positive definite correlation matrix. As discussed earlier, due to computational complexity the multivariate integration approach is applied only in the scenario with fewer subgroups (60 patients per arm within each subgroup). For the Bonferroni approach we calculate the adjusted p-values for each subgroup based on a univariate t-distribution with $N_j-1$ degrees of freedom (instead of a univariate normal distribution) to achieve a better finite sample approximation for subgroups of small sample size.

In addition to the approaches based on pseudo-outcomes, we consider an approach using the simple differences in group means. Let $\bar{Y}_{a,j}$ denote the mean and $n_{a,j}$ the number of patients on treatment $a$ in subgroup $\mathcal{S}_j$, and let $\bar{Y}_{a}$ and $n_a$ denote the corresponding overall quantities. The test statistic in subgroup $\mathcal{S}_j$ is given by $T_j=\frac{\bar{Y}_{1,j}-\bar{Y}_{0,j}-(\bar{Y}_{1}-\bar{Y}_{0})}{s_j}$, where $s_j$ allows for heteroskedasticity, using individual variance estimates based on the squared residual from the arm-specific mean within each treatment arm. P-values for the maximum statistic are approximated using the Bonferroni method utilizing a t-distribution based on the Satterthwaite approximation for the degrees of freedom for each subgroup.

In addition we run all methods in this section assuming homoskedasticity of the pseudo-outcomes, as described in Section \ref{ssec:homoskedastic}. In the truly homoskedastic scenarios this allows us to evaluate the loss of efficiency via estimation of a heteroskedastic variance. One difference is that for the homoskedastic Bonferroni methods normal quantiles (instead of t-quantiles) are used. 

\subsection{Performance measures}
\label{sec:performance_measure}

As performance measures, we consider the repeated sampling distribution of the p-value. Specifically, this is the probability approximated by each method that $T_{\max}$ exceeds the observed value of the maximum statistic.
\begin{itemize}
    \item Under homogeneity, we present the empirical distribution function of the obtained p-values, which should closely approximate the uniform distribution on the unit interval, $U([0,1])$, if they constitute valid p-values \citep{greenland2019valid}.
    \item Under heterogeneity, we display density plots of the p-values, together with the proportion of p-values less than 0.1 to evaluate the ability of each method to detect departures from homogeneity.
\end{itemize}

\subsection{Results}
\label{sec:results}

\begin{figure}
    \centering
    \includegraphics[width=0.75\linewidth]{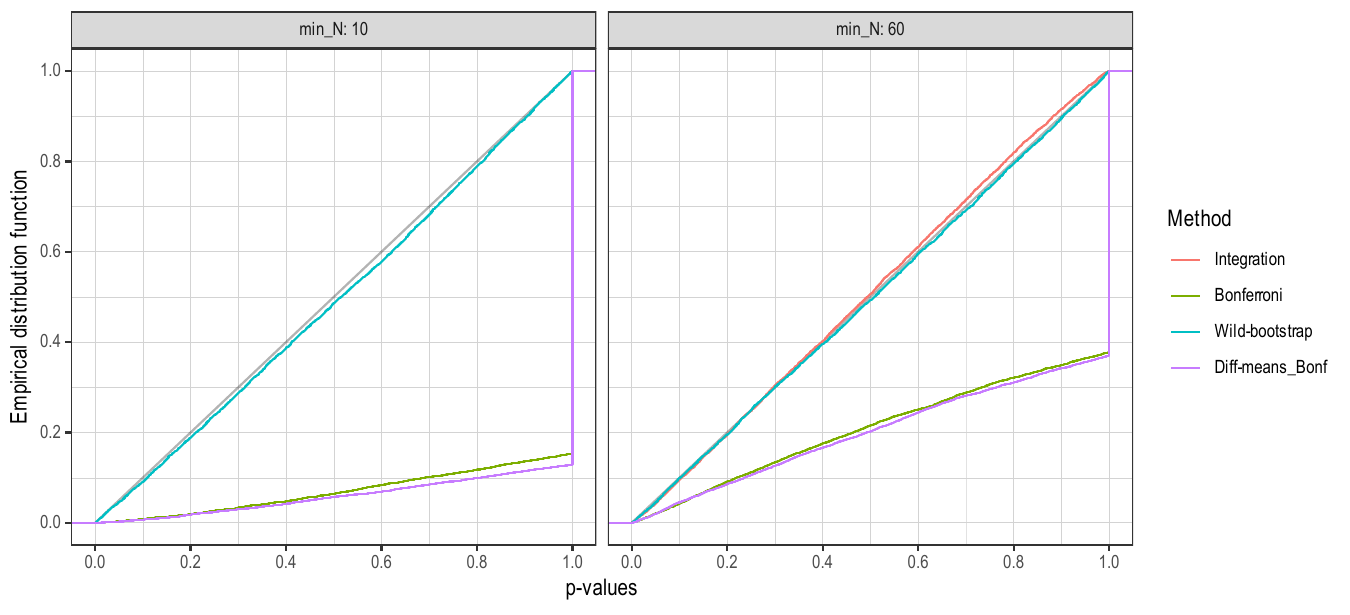}
    \caption{Empirical distribution function of p-values under no treatment effect heterogeneity. Results are pooled across the four scenarios from Table \ref{tab:sim_models}, thus resulting in 8000 simulations.}
    \label{fig:ecdf_ho}
\end{figure}

Simulations were performed with 2000 repetitions. Figure \ref{fig:ecdf_ho} displays the empirical distribution function of the p-values, along with the diagonal line corresponding to the $U([0,1])$ distribution. For scenarios with at least 10 observations in both treatment and control arms, the wild bootstrap approach closely follows the uniform distribution. The two Bonferroni-type approaches lie consistently below the diagonal (hence conservative) and do not follow the uniform distribution, showing a spike at 1. This is undesirable, as it makes the resulting p-values more difficult to interpret. The results for 60 observations per arm are similar; this case also includes the integration-based approach, which is essentially indistinguishable from the wild bootstrap approach in the lower part of the distribution.

\begin{figure}
    \centering
    \includegraphics[width=1\linewidth]{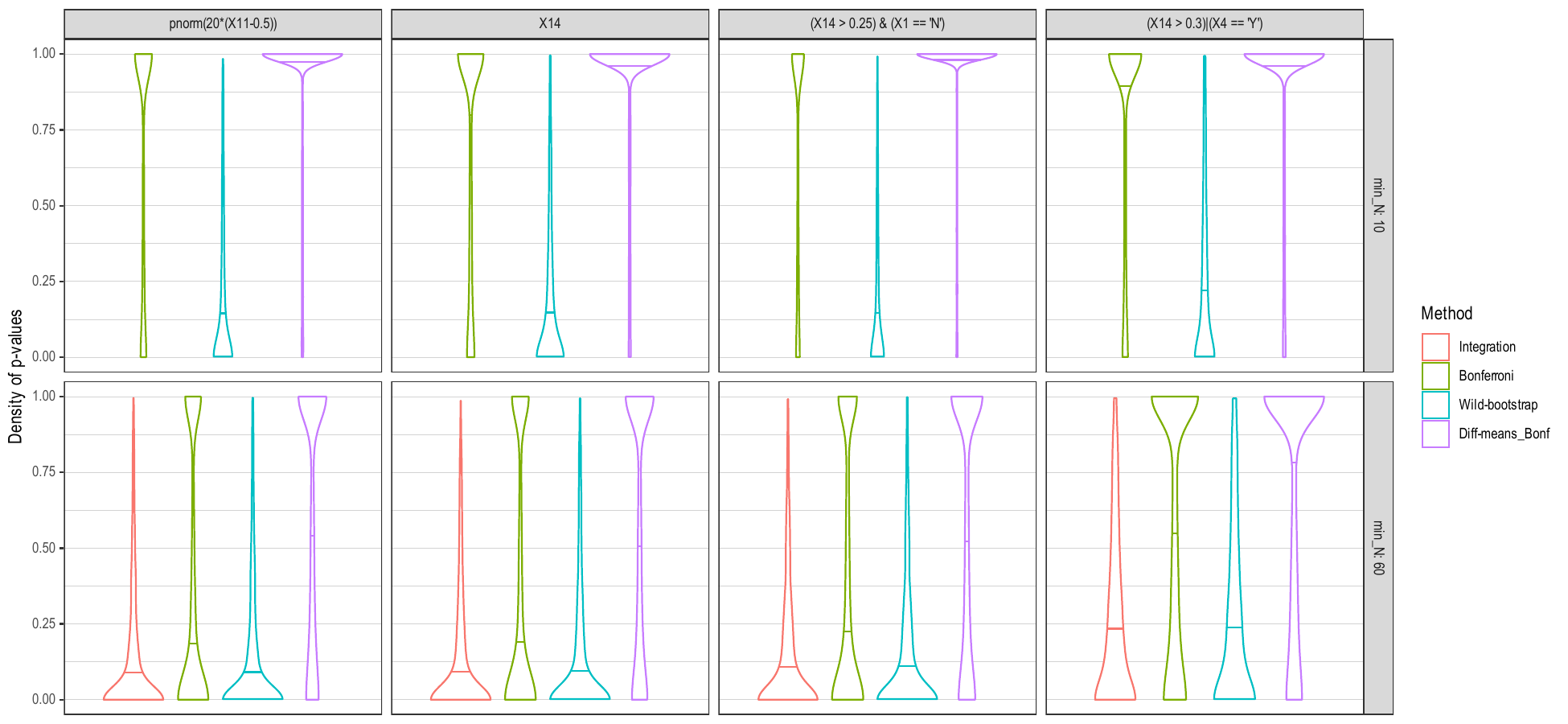}
    \caption{Distribution of p-values under treatment effect heterogeneity, shown as density plots. Horizontal lines indicate the median.}
    \label{fig:pval}
\end{figure}

Figure \ref{fig:pval} shows vertical density plots of the distribution of p-values under the scenario with $\beta_1=2\beta_1^*$, where treatment effect heterogeneity is present. The densities under the wild bootstrap and integration-based approaches are very similar. Compared to the Bonferroni-based approaches, the wild bootstrap and integration-based methods yield smaller median p-values (indicated by horizontal lines) and are generally shifted toward lower values. Between the two Bonferroni-based methods, the version using pseudo-outcomes performs slightly better, likely because the construction of pseudo-outcomes implicitly incorporates covariate adjustment.

\begin{figure}
    \centering
    \includegraphics[width=1\linewidth]{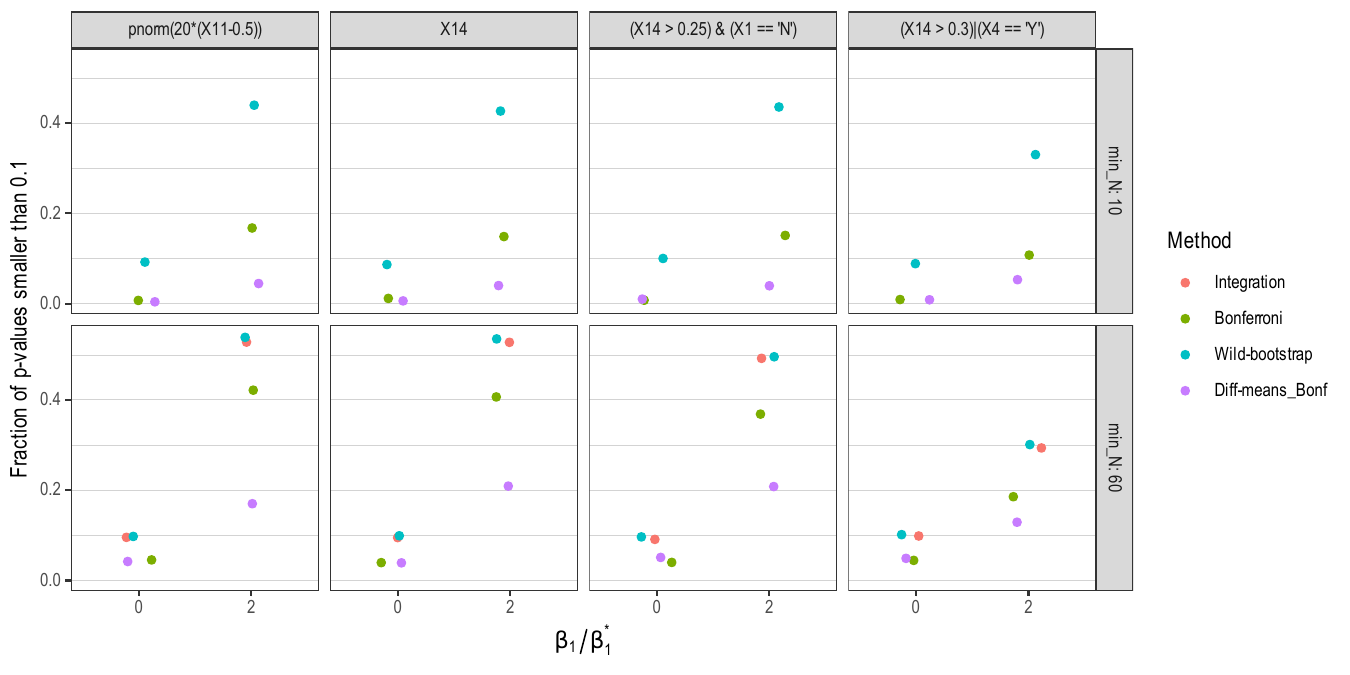}
    \caption{Proportion of p-values less than 0.1 across all scenarios.}
    \label{fig:power}
\end{figure}

Figure \ref{fig:power} summarizes the proportion of p-values less than 0.1, where 0.1 is used as an illustrative threshold. Under homogeneity, all approaches yield probabilities of about 0.1 or less, as expected from Figure~\ref{fig:ecdf_ho}. Under heterogeneity, the wild bootstrap approach outperforms the Bonferroni-based methods in the setting with at least 10 patients per arm and subgroup. For the setting with at least 60 patients per arm and subgroup, all methods show increased probabilities, likely due to reduced multiplicity when fewer subgroups are investigated. Overall, the results confirm that the integration- and bootstrap-based approaches perform similarly and better than the Bonferroni-based methods. Among the Bonferroni-based approaches, the version based on pseudo-outcomes consistently outperforms the one based on simple subgroup means. In Figures \ref{fig:ecdf_h0_app}, \ref{fig:power_app} in Appendix \ref{sec:app_plots} show the empirical distribution functions and detection plots stratified by each scenario, including the heteroskedastic scenario and the methods assuming homoskedasticity. The heteroskedastic wild bootstrap and integration methods achieve performance similar to the corresponding methods that assume homoskedasticity. In the heteroskedastic scenario with at least 10 patients per arm, the homoskedastic Bonferroni approaches are strongly anti-conservative, whereas the wild bootstrap remains calibrated also with the homoskedastic variance standardization. This is due to the fact that the wild bootstrap seems to approximate the correct heteroskedastic sampling distribution (see Section \ref{sec:meth}). In the heteroskedastic scenario with at least 60 patients per arm, however, the homoskedastic wild bootstrap and integration approaches seem approximately calibrated, with the Bonferroni methods being conservative. Interestingly, the heteroskedastic integration approach is slightly anti-conservative in this scenario, however only for the upper quantiles. 

\section{Example revisited}
\label{sec:example_rev}

In this section, we revisit the example from Section \ref{sec:example}. The pseudo-outcomes for the treatment effect were generated using an ensemble approach with the R package \emph{SuperLearner}. For each treatment arm, we applied 10-fold cross-fitting with LASSO and random forest as base learners. The resulting out-of-fold predictions for $\mu_0(X_i)$ and $\mu_1(X_i)$, together with the true propensity scores, were used in the equation for the pseudo-outcomes. Here the wild bootstrap approach described in Section~\ref{sec:meth} was then used to calculate the p-values.

\begin{figure}[htbp]
    \centering

    \begin{subfigure}{\textwidth}
        \centering
        \includegraphics[width=5in]{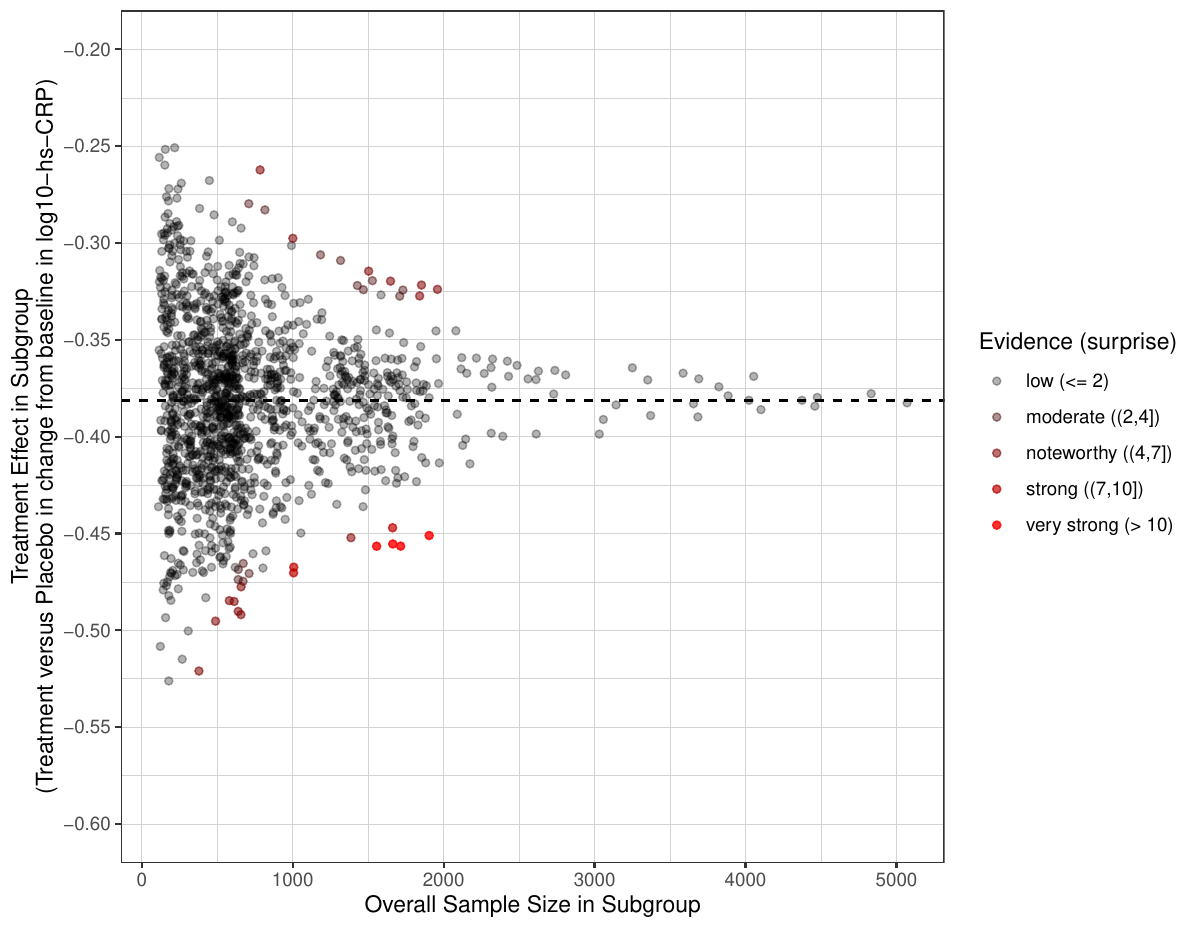}
        \caption{}
        \label{fig:funnel_plot_with_evidence}
    \end{subfigure}

    \vspace{0.5em}

    \begin{subfigure}{\textwidth}
        \centering
        \includegraphics[width=5in]{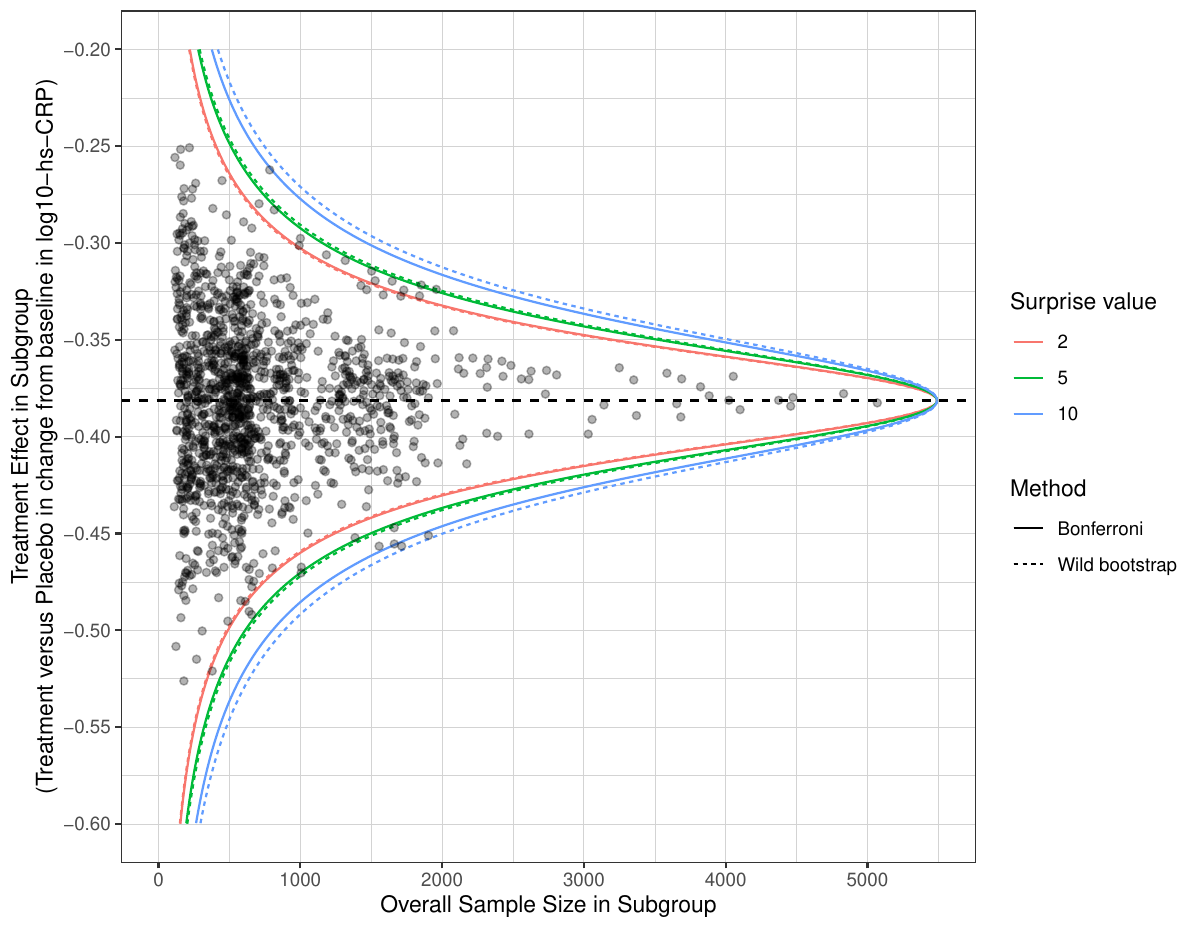}
        \caption{}
        \label{fig:funnel_with_lines}
    \end{subfigure}

\caption{Exhaustive subgroup treatment effect plots showing 1408 one-level or two-level subgroup treatment effects.
(a) Coloring of the subgroup dots corresponds to the level of evidence against homogeneity according to the reference distribution of the maximum statistic. The categories ``low'', ``moderate'', ``noteworthy'', ``strong'', and ``very strong'' correspond to surprise values of $\leq 2$, $(2,4]$, $(4,7]$, $(7,10]$, and $>10$, respectively.
(b) $\gamma$-homogeneity regions based on the wild bootstrap (dashed) and the Bonferroni approximation (solid) are shown for surprise values $S=2,5,$ and $10$, which correspond approximately to $\gamma=0.75, 0.97,$ and $0.999$, respectively. The dashed line denotes the overall treatment effect estimate.}    
\label{fig:funnel_plots}
\end{figure}

Figure~\ref{fig:funnel_plots}(a) shows the plot with the subgroups colored according to the corresponding evidence against homogeneity as determined by the underlying p-value. We measured this evidence in terms of surprise values \citep{cole2020,Greenland2023}. A surprise value (or S-value) is defined as $S = -\log_2(p)$ and can be interpreted as the amount of information, measured in bits, against the homogeneity assumption. For example, $S=1$ corresponds to one bit of information, equivalent to the evidence provided by a single fair coin toss coming up heads; $S=10$ corresponds to about ten bits of information, which is substantially stronger evidence. Larger S-values therefore indicate results that are more surprising under the homogeneity model and thus stronger evidence against it. Because they are computed from the multiplicity-adjusted $p_j$ introduced in Section~\ref{sec:meth}, these surprise values are descriptive measures of how extreme each subgroup is under an assumption of treatment effect homogeneity on all subgroups while accounting for the underlying multiplicity problem. The categories for the surprise value used in the plot are "low", "moderate", "noteworthy", "strong" and "very strong" corresponding to surprise values of $\leq2$, $(2,4]$, $(4,7]$, $(7,10]$ and $>10$\cite{sechidis2025watch}.

Some subgroups in Figure~\ref{fig:funnel_plots}(a) indicate strong evidence of treatment effect heterogeneity. Table~\ref{tab:subgroups} lists the five subgroups with the largest absolute maximum statistics in the simulated data. It is interesting to note that one subgroup (BMI $>$ 32) is part of all five subgroups. Note that a subgroup covering a large majority of patients may appear in combination with a subgroup such as BMI $>$ 32, but this does not mean that it is the driver of the differential treatment effect. In practice, findings such as those in Table~\ref{tab:subgroups} would need to be examined for clinical relevance and in the context of external evidence (e.g., whether independent data exist that corroborate these results) and likely trigger further discussions. 

\begin{table}[ht]
\caption{Five subgroups with the largest absolute test statistics}
\label{tab:subgroups}
\centering
\begin{tabular}{lrrrr}
  \hline
Subgroup & N (trt) & N (ctrl) & Trt. effect & $|T|$\\ 
  \hline
 No GOUT \& BMI $>$ 32                            & 679 & 1036 & -0.46 & 5.52\\ 
 BMI $>$ 32                                       & 767 & 1137 & -0.45 & 5.41 \\ 
 RACE $=$ "White" \&  BMI $>$ 32                   & 675 & 988 & -0.46 & 5.31 \\ 
 BMI $>$ 32 \& ETHNIC $=$ "Not hispanic or latino" & 622 & 934 & -0.46 & 5.14 \\ 
 Aspirin \& BMI $>$ 32                            & 668 & 994 & -0.45 & 4.68 \\
   \hline
\end{tabular}
\end{table}

If one is willing to assume homoskedastic variance of the pseudo-outcomes, then one can construct $\gamma$-homogeneity regions as described in Section~\ref{ssec:homoskedastic}. 
Figure~\ref{fig:funnel_plots}(b) shows the intervals derived from the wild bootstrap and Bonferroni approaches, which are well aligned. It is surprising to see that the conservative Bonferroni approach leads to narrower intervals for large $\gamma$. We attribute this to the limitations of the normal approximation used for the Bonferroni approximation for the homoskedastic setting: for extreme critical values (e.g., a surprise value of 10), the normal tails may underestimate the empirical tails of the test statistic distribution, particularly when small subgroups are included in the analysis. Note that from the simulation results we know that the wild bootstrap can be expected to work appropriately even when a homoskedastic variance estimate is used, as discussed in Section \ref{sec:results}. Note that Table~\ref{tab:subgroups_homo} in Appendix \ref{sec:app_tbl} contains the test statistics under homoskedasticity, providing largely similar results as in the heteroskedastic setting.

\section{Discussion}
\label{sec:discussion}

Exploration of treatment effect heterogeneity is a critical component in clinical trials, as it may suggest subgroups of patients who benefit more or less from treatment. Exhaustive subgroup treatment effect plots provide an intuitive and comprehensive visualization for such exploratory analyses. In this paper, we developed a computationally feasible method to assess the evidence against homogeneity for treatment effect differences, enabling systematic and interpretable summaries across large numbers of subgroups. This approach balances the need for rigorous statistical assessments in exploratory settings by considering how surprising the observed heterogeneity of certain subgroups is under a model of homogeneous treatment effects. 

We interpret p-values as divergence p-values\citep{Greenland2023}, which quantify the degree of incompatibility between the observed data and what would be expected under a model of no treatment effect heterogeneity. Unlike conventional interpretations of decision p-values, which are tied to thresholds for binary decisions, divergence p-values serve as graded measures of evidence. They can be further expressed as surprise values. For practical applications, research teams may find surprise values useful for communication, as they translate the strength of evidence into an intuitive scale that avoids the pitfalls of binary dichotomization.

The R package \emph{subscreen} provides an interactive graphical user interface for subgroup analysis \cite{muysers2020systematic, kirsch2022subscreen}.  Such interactive displays allow users either to identify points corresponding to specific subgroups of interest or to click on points to determine which subgroup they represent. This functionality is valuable for obtaining an overall impression of how treatment effects vary with baseline covariates. A natural extension would be to enhance such interfaces with functionality for displaying  surprise values as introduced in this paper, thereby combining interactivity with statistical calibration.

We compared several methods for assessing evidence against homogeneity in this paper. The approaches using numerical integration or wild bootstrap based on pseudo-outcomes for the treatment effect perform well across the scenarios considered. In particular the wild bootstrap approach lead to uniformly distributed p-values under the scenario of homogeneity. The resulting p-values are hence easier to interpret than, for example, those from Bonferroni-based approaches. The wild bootstrap approach is computationally more expensive than Bonferroni but less involved than the integration approach. Overall, the wild bootstrap approach hence provides the most convincing performance in our simulations.

In this paper we utilized the difference in conditional means as the treatment effect measure of interest. Whether or not treatment effect heterogeneity exists and the extent of such heterogeneity depend on the chosen treatment effect measure; see, for example, \cite{chen2025comparing} for a discussion. An extension of the proposed method to alternative treatment effect measures may be considered in future work.

Our inferential framework is conditional on the realized baseline covariates $X_1,\ldots,X_N$. Consequently, subgroup definitions, memberships, and sizes are treated as fixed once the baseline covariates have been observed. This framework also accommodates data-dependent subgroup definitions, such as empirical tertiles of continuous baseline covariates, provided that their construction depends only on baseline covariate information and not on outcomes. 

The proposed approach can also be used within the Workflow for Assessing Treatment effeCt Heterogeneity (WATCH) framework introduced by \cite{sechidis2025watch}. This framework emphasizes three principles: the integration of external evidence to evaluate the plausibility and relevance of subgroup findings; the use of global heterogeneity tests to assess overall departures from homogeneity; and the reliance on exploratory displays to investigate potential drivers of heterogeneity. In addition, interactive and user-friendly visualizations, such as exhaustive subgroup treatment effect plots displaying levels of evidence as discussed above, could further support decision-making by allowing stakeholders to explore subgroup patterns while simultaneously assessing their robustness against multiplicity adjustments.

%\backmatter

\noindent \textbf{Acknowledgments}\\
We thank Professor Oliver Dukes for helpful comments and discussions on an earlier version of the paper. An AI assistant (Claude Opus 5.5) was used during the revision to check derivations, to suggest methodological alternatives (including the wild bootstrap approach), to implement changes in the R code, and to draft and edit parts of the text. The authors reviewed and verified all output and take full responsibility for the content.
%\bmsection*{Financial disclosure}

%None reported.

%\bmsection*{Conflict of interest}

%The authors declare no potential conflict of interests.

\bibliographystyle{abbrv}
\bibliography{reference_lib_revision.bib}

\appendix

\section{Covariate data description for simulated, synthetic data}
\label{sec:app_bl}

\begin{table}[h]
\caption{Naming of baseline variables and used categories}
\label{tbl:variables}
\tiny{
\centering
\begin{tabular}{lll}
  \hline
Names & Description & Categories \\ 
  \hline
AGE &  & (22,56], (56,66], (66,110] \\ 
  ASPRNFL & Baseline Aspirin Therapy & N, Y \\ 
  BASECRP &  & (-0.72,0.5], (0.5,0.75], (0.75,2.5] \\ 
  ETHNIC &  & HISPANIC OR LATINO, NOT HISPANIC OR LATINO, UNKNOWN \\ 
  GLYCEM & Glycemic Status at Baseline & Diabetic, Normoglycemic, Prediabetic \\ 
  LBEGFG1B & Baseline EGF rate (mL/min) & $<$ 60 mL/min/SA, $>$= 60 to $<$ 90 mL/min/SA, $>$= 90 mL/min/SA \\ 
  LBLOGHDL & Base log value of HDL cholest. (mmol/L) & (-0.89,0.0086], (0.0086,0.1], (0.1,0.59] \\ 
  LBLOGLDL & Base log value of LDL-C derived (mmol/L) & (-1.7,0.26], (0.26,0.4], (0.4,1.2] \\ 
  LBLOGTRI & Base log value of Triglycerides (mmol/L) & (-0.62,0.11], (0.11,0.29], (0.29,1.7] \\ 
  MHGOUTFL & Medical History of Gout & N, Y \\ 
  QMITGR3 & Time Since Qualifying MI & $<$ 12 months, $>$= 12 months \\ 
  RACE &  & ASIAN, OTHER, WHITE \\ 
  REGION1 &  & ASIA, CENTRAL EUROPE, LATIN AMERICA, NORTH AMERICA, \\ 
          &  & OTHERS, WESTERN EUROPE \\ 
  SEX &  & F, M \\ 
  SMOKE & Smoking Status at Baseline & Current smoker, Former smoker, Never \\ 
  STATINB & Base median daily dose of statin (mg) & High Dose, Low Dose, Medium Dose, No Dose \\ 
  VSBMIS & Screening BMI (kg/m2) & (11,28], (28,32], (32,94] \\ 
  VSSTDBMB & Baseline mean sitting diastolic BP (mmHg) & (32,74], (74,82], (82,150] \\ 
  VSSTSBMB & Baseline mean sitting systolic BP (mmHg) & (120,140], (140,240], (63,120] \\ 
   \hline
\end{tabular}}
\end{table}

\newpage

\section{Unbiasedness of pseudo-outcomes in randomized setting}
\label{sec:app_dr}

The expectation of the pseudo-outcome $E(\widehat{\varphi}_i|X_i)$ is given by $E(\widehat{\mu}_1(X_i) - \widehat{\mu}_0(X_i) + \frac{A_i - \pi}{\pi(1 - \pi)} \left(Y_i - \widehat{\mu}_{A_i}(X_i) \right)|X_i)$.

Here $\pi$ is the known randomization proportion. In addition because of the use of cross-fitting $\widehat{\mu}_1(X_i)$ and $\widehat{\mu}_0(X_i)$ are fixed constants and using the law of iterated expectation we have that 

\begin{eqnarray}
  E(\widehat{\varphi}_i|X_i)&=& (\widehat{\mu}_1(X_i) - \widehat{\mu}_0(X_i))+\pi\frac{1}{\pi}(E(Y_i|A_i=1,X_i)-\widehat{\mu}_1(X_i))-(1-\pi)\frac{1}{1-\pi}(E(Y_i|A_i=0,X_i) - \widehat{\mu}_0(X_i))  \nonumber \\
                           &=& \widehat{\mu}_1(X_i) - \widehat{\mu}_0(X_i) + \mu_1(X_i)-\widehat{\mu}_1(X_i)-\mu_0(X_i)+\widehat{\mu}_0(X_i)  \nonumber \\
  &=& \mu_1(X_i)-\mu_0(X_i) \nonumber
\end{eqnarray}

\newpage

\section{Additional plot}
\label{sec:app_plots}

\begin{figure}[h]
    \centering
    \includegraphics[width=1\linewidth]{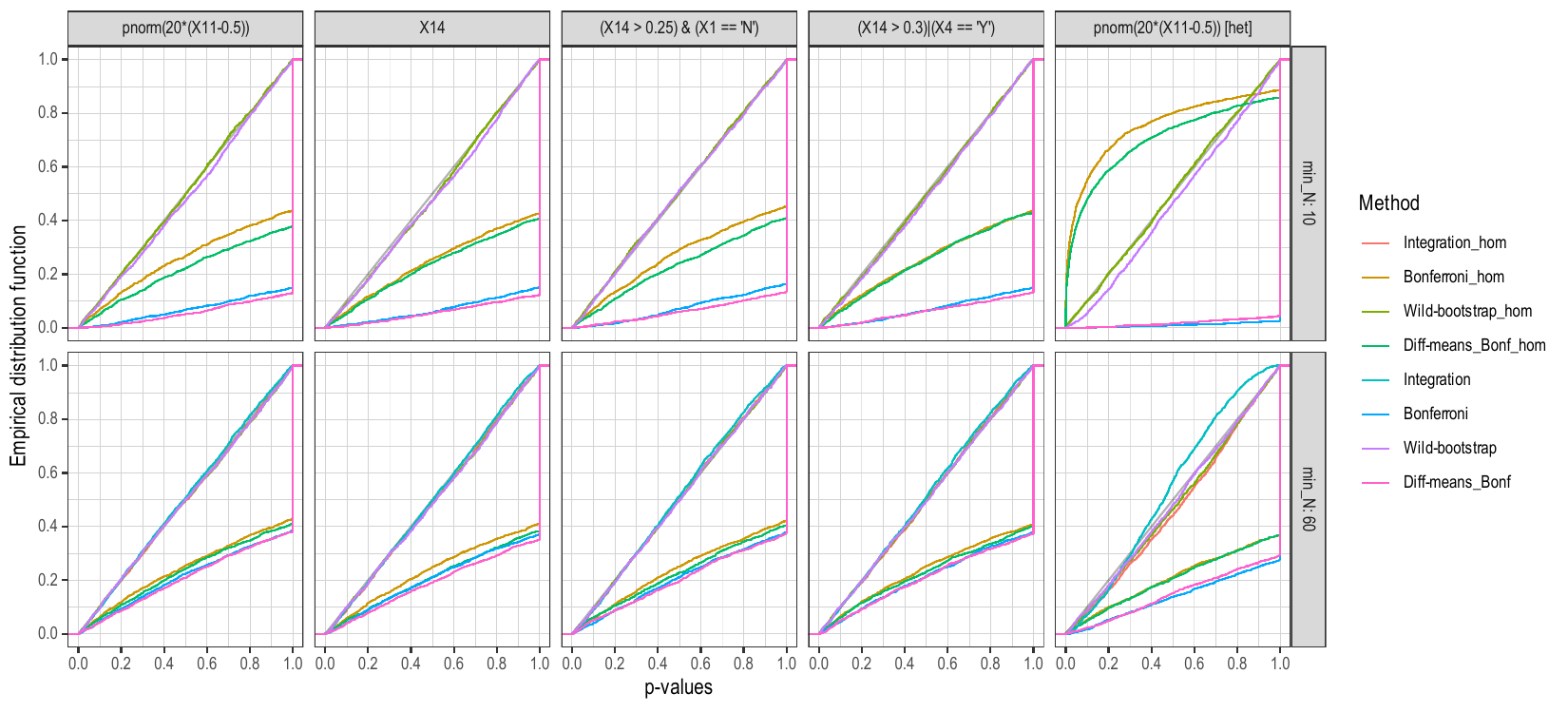}
    \caption{Empirical distribution function of p-values under no treatment effect heterogeneity. Now also a heteroskedastic scenario is included as well as methods that assume homoskedasticity.}
    \label{fig:ecdf_h0_app}
\end{figure}

\begin{figure}[h]
    \centering
    \includegraphics[width=1\linewidth]{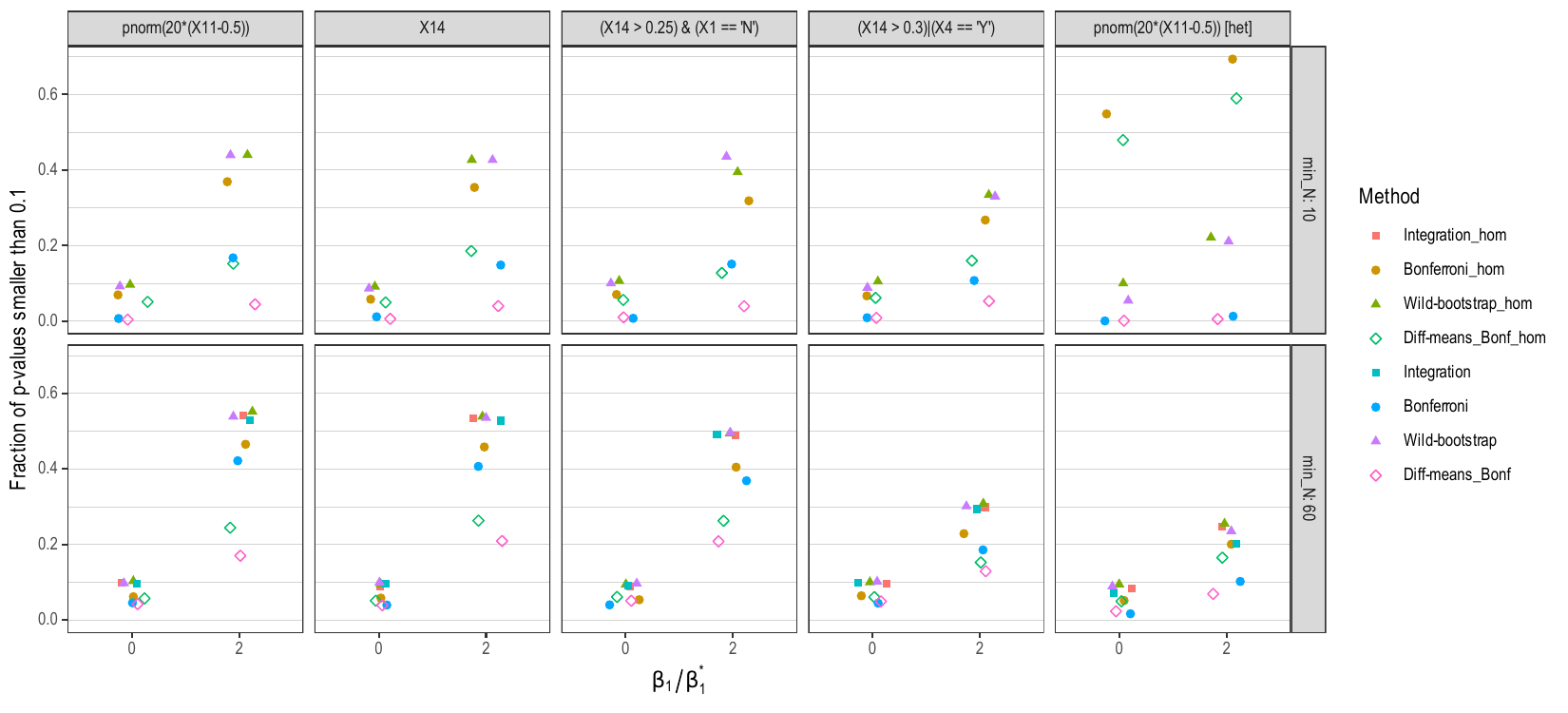}
    \caption{Proportion of p-values less than 0.1 across all scenarios. Now also a heteroskedastic scenario is included as well as methods that assume homoskedasticity. }
    \label{fig:power_app}
\end{figure}

\newpage

\section{Additional table for simulated, synthetic data}
\label{sec:app_tbl}

\begin{table}[ht]
\caption{Five subgroups with the largest absolute test statistics (under a homoskedasticity assumption)}
\label{tab:subgroups_homo}
\centering
\begin{tabular}{lrrrr}
  \hline
Subgroup & N (trt) & N (ctrl) & Trt. effect & $|T|$\\ 
  \hline
 BMI $>$ 32                                       & 767 & 1137 & -0.45 & 5.13 \\ 
 No GOUT \& BMI $>$ 32                            & 679 & 1036 & -0.46 & 5.12\\ 
 RACE $=$ "White" \&  BMI $>$ 32                   & 675 & 988 & -0.46 & 4.93 \\ 
 BMI $<28$ \& LogHDL $>$ 0.1                       & 314 & 470 & -0.26 & 4.91 \\    
 BMI $>$ 32 \& ETHNIC $=$ "Not hispanic or latino" & 622 & 934 & -0.46 & 4.78 \\ 
   \hline
\end{tabular}
\end{table}

\end{document}